\documentstyle[preprint,aps,tighten]{revtex}
\newcommand{\beq}{\begin{equation}}
\newcommand{\eeq}{\end{equation}}
\newcommand{\beqa}{\begin{eqnarray}}
\newcommand{\eeqa}{\end{eqnarray}}

\newcommand{\nm}{$\nu_{\mu}$}
\newcommand{\ns}{$\nu_s$}
\newcommand{\nue}{$\nu_e$}

\def\sm{standard model}

\def\ifmath#1{\relax\ifmmode #1\else $#1$\fi}
\def\half{\ifmath{{\textstyle{1 \over 2}}}}
\def\plb#1{Phys.\ Lett.\ {\bf B#1}}
\def\prd#1{Phys.\ Rev.\ {\bf D#1}}

\def\zpc#1{Z.~Phys.\ {\bf C#1}}

\renewcommand{\thefootnote}{\fnsymbol{footnote}}

\draft
\begin{document}
{\tighten

\preprint{
\vbox{
      \hbox{SLAC-PUB-7888}
      \hbox{hep-ph/9807511}
      \hbox{July 1998} }}
 
\title{
Atmospheric $\nu_{\mu}$ Deficit from Decoherence
\footnotetext{Research supported
by the Department of Energy under contract DE-AC03-76SF00515}}
\author{Yuval Grossman and  Mihir P. Worah}
\address{ \vbox{\vskip 0.true cm}
Stanford Linear Accelerator Center \\
        Stanford University, Stanford, CA 94309}
\maketitle

\begin{abstract}%
The simplest explanation for the observed deficit of atmospheric
muon neutrinos is that they have oscillated into tau or sterile
neutrinos with an oscillation length of the order of the Earth
diameter. In order to confirm this hypothesis, 
all other possible explanations
should be ruled out. We propose that a viable alternative hypothesis
is that the muon neutrino deficit is caused by flavor sampling events
that result in a loss of coherence. The coherence length of the muon
neutrinos is expected to be approximately one Earth diameter. We give
predictions and experimental tests of this scenario.
\end{abstract}

}%end of tighten
\newpage
\renewcommand{\thefootnote}{\alph{footnote}}

The recent data on the flux of atmospheric neutrinos reported by the  
Super-Kamiokande (SuperK) collaboration gives 
strong evidence for neutrino mass \cite{SuperK}.
The old data that indicated that the integrated  
ratio of muon like events to electron like
events is smaller than expected was confirmed. 
Moreover, this ratio was shown to depend on the zenith angle and 
on the energy of the incident neutrinos. 
In particular, no evidence for disappearance 
was found for down coming events, 
where the neutrinos travel short distances.
While for up coming events, 
where the neutrino travel much longer, 
the number of muon events was about half of the expected number.
Furthermore, a depletion in the expected number of muon
events as a function of $(L/E)$, where $L$ and $E$ are the inferred
flight length and energy of the neutrino, was found.

The simplest explanation for the disappearance of the muon neutrino
(\nm) is that it has oscillated into a tau neutrino ($\nu_{\tau}$)
or a sterile neutrino (\ns) with an
oscillation length which is comparable to the Earth diameter.
The best fit to the data indicates mixing with
a mass squared difference $\Delta m^2 =O(10^{-3})\,$eV$^2$, and a
vacuum mixing angle $\sin^2 2\theta = 1$. 
There exist other less likely explanations for the \nm\ deficit
that need to be ruled out based on the data itself 
in order to undoubtedly confirm
the simple and elegant oscillation hypothesis. 
One is that the \nm\ decays between its production in the atmosphere
and its detection on the earth \cite{No_Decay}. 
Another is that a violation of Lorentz invariance or 
a breakdown of the equivalence principle results in \nm\ disapperance
\cite{Lor_inv}.

In this note we postulate another speculative mechanism that seems to
agree with the data. 
We propose that the \nm\ coherence length is somewhat less than the 
diameter of the Earth. This coherence loss is due to some
unknown mechanism that measures the neutrino flavor.\footnote{Coherence 
loss due to wave packet separation
does not measure the flavor of the propagating
state and is irrelevant here.}
As a simple example consider the case of \nm-\ns\ mixing. In
this scenario,  the down going \nm's travel a distance much
shorter than their coherence length, and hence remain
\nm's. The up going \nm's, however, travel a distance of a
few times their coherence length, and hence are reduced to the
equilibrium  state containing an equal amount of \nm\ and
\ns. Since, at present, there is no evidence that flavor remains
coherent over distance scales larger than a few kilometers, 
it is important to rule out
this mechanism in order to establish the neutrino oscillations
hypothesis and its parameters.

With \sm\ interactions
the coherence length for neutrinos of a few
GeV energy is several orders of magnitude 
larger than the earth diameter. Therefore, in order for
decoherence to explain the atmospheric neutrino deficit one needs a
new mechanism of flavor dependent coherence loss for the
neutrinos. (It must be flavor dependent in order to leave the 
atmospheric electron neutrinos unaffected.)
Some examples of scenarios that could result in 
such a decoherence are a very large 
neutrino background \cite{MoNu}, flavor dependent interaction into
an extra dimension \cite{Nima}, or even small violations of quantum 
mechanics. We do not explore any of these ideas in detail since our
purpose is to concentrate on the general features
of the decoherence explanation and, in particular, on its experimental
tests.

Neutrino propagation with coherence loss for the two species case
can be described by the equation \cite{Stodolsky}:
\beq
\frac{d{\bf P}}{dt} = {\bf v} \times {\bf P} - D {\bf P}_T\,.
\label{propogation}
\eeq
Here ${\bf P}$ is a ``polarization'' vector such that ${\bf P}_z=+1(-1)$ 
corresponds
to a pure \nm\ (\ns). ${\bf P}_T$ is the transverse part of ${\bf P}$,
${\bf v}$ is the mass eigenstate vector such that 
$|{\bf v}|=\Delta m^2/2E$, and
it is an angle $2\theta$ from the $z$ axis, and
$D$ is the (energy dependent) damping coefficient \cite{Stodolsky}. 
While the general
solution to the above formula is complicated, we can understand its
implications in  several limiting cases. The weak damping limit, 
$D \ll |{\bf v}|$, corresponds to the large $\Delta m^2$ limit where the
oscillations are too rapid to be observed, and must be averaged
over. In this case the \nm\ survival probability is given by 
\beq
P_{\mu\mu}(t) = \frac{1}{2}\left[1 + \cos 2\theta e^{-t/\tau(E)}\right]\,,
\qquad \tau(E)^{-1} = D(E)\sin2\theta\sqrt{1-{\sin^22\theta \over 4}}\,.
\label{survival}
\eeq
In the case of critical
damping ($D \sim |{\bf v}|$), the damping time is similar as above, however
the oscillations will not average out, and the survival probability
will be more complicated. Finally in the over damped case, the
relaxation time is much longer \cite{Stodolsky} and is not of relevance
here.

We concentrate on the weakly damped case given by
Eq.$\,$(\ref{survival}) and propose that this is the form the muon
survival probability at SuperK should be fitted for. 
While we cannot perform a fit of Eq. (\ref{survival})
to the data, it seems it can agree with it.
The mean free path needed to explain the SuperK data should
be somewhat smaller than the Earth's diameter, $\tau \sim
10^{-2}\,$s,
where $\tau$ is presumably a smooth function of the energy.
Moreover, in order to not significantly deplete the downward going
neutrinos, we require $\sin2\theta \lesssim 0.4$.
Then, the suppression for the up going muons is
about a half, and it is minimal (less then 5\%) 
for the down going ones.
Note, that this implies $\Delta m^2 \sin 2\theta \gg 10^{-3}\,$eV$^2$
in order for the weak damping limit to be applicable.

The limits on this decoherence mechanism from terrestrial experiments
are rather weak. The existing experiments have a baseline of
no more than about 1 km. Hence Eq.$\,$(\ref{survival}) predicts  
that the total transition probability is less then $10^{-3}$ if
$\tau$ does not depend on energy.
This is consistent with the current bounds from disappearance
($P_{\mu \mu\!\!\!\slash} < 0.01$) and 
appearance ($P_{\mu\tau} < 0.002$) experiments \cite{pdg}.
Note that in some of the accelerator experiments the neutrino energies are
much larger than for the atmospheric neutrinos, 
e.g., $E \sim O(100)\,$GeV \cite{CNTR}. This seems to indicate that 
$\tau$ should not rapidly decrease with energy. Recall, however, that  
if the oscillation length is much longer than the baseline, 
Eq. (\ref{survival}) is not applicable and the
transition probability is much more suppressed. We thus conclude that 
a strong reduction of $\tau$ with energy is excluded for large $\Delta m^2$.
For example, from Ref. \cite{CNTR} we conclude that $\tau \propto E^{-n}$
with $n \gtrsim 1$ is excluded for $\Delta m^2 \gtrsim 50\,$eV$^2$.

Astrophysical constraints can also be used to check some of the 
general features of the mechanism we have proposed.
Big Bang Nucleosynthesis (BBN) constraints disfavor \nm-\ns\ mixing.
To explain the atmospheric neutrino data we require the \ns\
interaction time to be much shorter than that for the standard
neutrinos. If this continued to hold unchanged 
in the early universe, it would cause 
the \ns\ to be thermally populated at the epoch of BBN
which is disfavored by the data \cite{BBNdata}.
However, we cannot rule this possibility out
since the \ns\ may have exotic interactions that result in a
tiny effective mixing angle in the early universe,
and moreover the energy dependence of $\tau$ is unknown.

The Solar neutrinos are primarily electron neutrinos (\nue) and hence 
are not directly related to our case.  
If, however,  the \nue\ undergo a similar mixing with \ns,
then a decoherence length significantly below $1\,$AU 
would lead to an energy
independent suppression of the solar neutrino flux in contradiction
with the data \cite{BKS}. Since the solar neutrinos have energies in the MeV
range (compared to $E \sim 1\,$GeV for the atmospheric neutrino)
we can conclude that either $\tau$ decreases with energy, or
that the \nue - \ns\ mixing is insignificant.

The best way to rule out our proposal is to observe an oscillation
pattern in  $(L/E)$ at SuperK.  At present, the
uncertainties in $(L/E)$ at SuperK are large resulting in a smearing
out of any possible oscillation pattern.
Even with the smearing, the disappearance probability
as a function of $E$ for the two cases 
may be different depending on the energy dependence 
of $\tau$. 
For the oscillation scenario, the disappearance 
probability goes as $E^{-1}$ after smearing. If, for example,  
$\tau$ decreases with energy as indicated by
our discussion of solar neutrinos, the disappearance 
probability will increase with $E$.

The up coming long
baseline neutrino oscillation experiments will be sensitive to  
much of the preferred parameter space 
of the decoherence hypothesis. From our considerations earlier in the
paper, the preferred range for the mixing parameters is 
$0.01\,$eV$^2 \lesssim \Delta m^2 \lesssim 10\,$eV$^2$ and 
$0.01 \lesssim \sin 2\theta \lesssim 0.4$ where the smaller values for
$\Delta m^2$ are correlated with larger values for $\sin 2\theta$. 
Thus, the signal 
will be different from that predicted by the oscillation
explanation of the SuperK data. If $\tau$ is independent of energy,
the disappearance will be due to a combination of decoherence and
oscillation. If, however, $\tau$ decreases with energy 
the decoherence effect
will be minimal and the signal will be purely due to oscillations.

An important prediction of our proposal is that for $t \gg \tau$,
$P_{\mu\mu} = \half$.  This is in contrast to the oscillation scenario
where averaging over many oscillation lengths results in 
$P_{\mu\mu} =1 - \half\sin^22\theta$ which could a priori have any value
from $\half$ to $1$. 
While $\tau$ can be fine tuned such that $P_{\mu\mu}$ for the atmospheric
neutrinos saturates at a value different from half, in general,
this will not be the case. If, when more data becomes available,
one finds that the saturation is to a value different from half, it
will disfavor the decoherence mechanism.

Before concluding let us remark that 
one can generalize this decoherence mechanism to 
massless neutrinos. The general requirements for the decoherence mechanism
to work are the presence of a mixing angle between the propagating
eigenstate and the ``flavor'' eigenstate -- the state that is measured;
and that the oscillation length is shorter than the decoherence length.
The first requirement can be met if the decoherence mechanism does not measure
the weak flavor but rather some superposition of the muon and sterile
neutrinos. The second requirement can be satisfied if the unknown
interactions generate an ``effective'' mass for one of the neutrinos.

To summarize, we propose that it is possible that 
the deficit in \nm\ observed at SuperK
is not due to the fact that \nm's oscillate over an Earth diameter,
but rather that they have oscillated many times and have lost
coherence due to some unknown mechanism over this distance.
While this explanation is admittedly speculative, it needs to be ruled
out in order for the oscillation hypothesis to be confirmed.

\bigskip
{\noindent \bf \it Acknowledgements.}
We thank N. Arkani-Hamed,
J. Learned, R. Plaga, G. Sigl and T. Weiler for helpful discussions.

%\nopagebreak

{\tighten

}


\begin{thebibliography}{}

\bibitem{SuperK}
The Super-Kamiokande Collaboration, Y. Fukuda et al., hep-ex/9807003.

\bibitem{No_Decay}
T.J. Weiler, private communication; 
V. Barger, J.G. Learned, S. Pakvasa and T.J. Weiler, in preparation.

\bibitem{Lor_inv}
S. Glashow, talk presented at Neutrino98.

\bibitem{MoNu}
Large local neutrino density was invoked for other purposes in 
R.N. Mohapatra and S. Nussinov, \plb{395}, (1997) 63.

\bibitem{Nima}
N. Arkani-Hamed, private communication.

\bibitem{Stodolsky}
L. Stodolsky, \prd{36} (1987) 2273;
G.G. Raffelt, ``Stars as laboratories for fundamental physics'',
Chicago Univ. Press, 1996.

\bibitem{pdg}
C. Caso {\it et al.} (Particle Data Group), 
European Physical Journal {\bf C3} (1998) 1.

\bibitem{CNTR}
I.E. Stockdale {\it et al.}, \zpc{27} (1985) 53.

\bibitem{BBNdata}
For a recent review see e.g., A.D. Dolgov, astro-ph/9807134.

\bibitem{BKS}
See, e.g., J.N. Bahcall, P.I. Krastev and A. Yu. Smirnov, hep-ph/9807216.

\end{thebibliography}
\end{document}